# Magnetic flux trapping in hydrogen-rich high-temperature superconductors

V. S. Minkov[1]*, V. Ksenofontov[1], S. L. Bud'ko[2,3], E. F. Talantsev[4,5], M. I. Eremets[1]*

[1]*Max Planck Institute for Chemistry; Hahn-Meitner-Weg 1, 55128 Mainz, Germany*

[2]*Ames Laboratory, U.S. Department of Energy, Iowa State University; Ames, IA 50011, United States*

[3]*Department of Physics and Astronomy, Iowa State University; Ames, IA 50011, United States*

[4]*M.N. Mikheev Institute of Metal Physics, Ural Branch of the Russian Academy of Sciences, S. Kovalevskoy St 18, 620108 Ekaterinburg, Russian Federation*

[5]*NANOTECH Centre, Ural Federal University, Mira St 19, 620002 Ekaterinburg, Russian Federation*

*Corresponding authors. Email: v.minkov@mpic.de (V.S.M); m.eremets@mpic.de (M.I.E.)

## ABSTRACT

Recent discoveries of superconductivity in various hydrides at high pressures have shown that a critical temperature of superconductivity $T_c$ can reach near-room-temperature values[1-3]. However, experimental studies are severely limited by high-pressure conditions, and electrical transport measurements have long been the primary technique for detecting superconductivity in hydrides. Herein, we dramatically changed the protocol for magnetic measurements and probed the trapped magnetic flux in two near room-temperature superconductors $H_3S$ and $LaH_{10}$. The specific behavior of the trapped flux generated under zero-field-cooled and filed-cooled conditions unequivocally proved superconductivity in these materials. In addition, we determined $T_c$, critical currents and their temperature dependence, the lower critical field $H_{c1}$, London penetration depth, full penetration field, pinning properties and flux creep. We revealed, that the absence of pronounced Meissner effect is associated with the very strong pinning of vortices inside samples. Contrary to traditional magnetic susceptibility measurements, magnetic response from the trapped flux is not affected by the background signal of the bulky diamond anvil cell due to the absence of external magnetic fields. This approach can be a powerful tool not only for a routine screening of new superconducting materials at high pressures, but also for studying multiphase samples or samples having a low superconducting fraction at ambient pressure.

## INTRODUCTION

Current search for room-temperature superconductivity has progressed due to advances in high-pressure studies. A decade ago, superconductivity was as assumed to be a solely low-temperature phenomenon with the highest $T_c$ of approximately 130 K found in mercury[4]- and thallium[5]-containing cuprate superconductors. The discovery of superconductivity in $H_3S$ with a $T_c$ of ~200 K[1] changed this paradigm and paved the way for higher $T_c$s in conventional superconductors for which BCS theory[6-8] was established. Thereafter, $T_c$s up to ~250 K were found in transition metal superhydrides $YH_9$[9] and $LaH_{10}$[10-12]. These experimental discoveries were anticipated by accurate theoretical predictions[13-15], which became feasible owing to the continuous development of ab-initio methods for prediction of crystal structure and computation of superconducting properties[16-18].

Although extreme pressure conditions render these record high-$T_c$ superconductors unsuitable for technological applications, the study of high-temperature superconductivity in various phases of hydrogen-rich compounds has highlighted the fundamental problem of the ultimate value of $T_c$ possible in a superconductor. Latest theoretical predictions suggest that very high multi-megabar pressures are required for superconductivity substantially above room temperature[19,20]. Another promising direction of high-pressure studies includes investigating effective methods for the stabilization of high-$T_c$ phases at lower pressure values, ideally at ambient conditions, for example by doping[21-23].

However, high pressures necessitate tiny micrometer-sized samples and drastically limit the available techniques for experimental study of superconductivity. When cooled below its critical temperature, a superconductor exhibits two characteristic properties: it abruptly loses electrical resistance and generates superconducting current loops that cancel the imposed magnetic field within its interior (Meissner effect). These two fundamental effects should be observed experimentally for the ultimate evidence of superconductivity in a newly discovered material.

Electrical transport measurements have been successfully adopted in the a multi-megabar pressure range and have become the primary experimental technique to detect superconductivity[24-26]. They provide data on $T_c$ and its dependence on pressure, and the estimates of the upper critical magnetic field $\mu_0 H_{c2}$ and coherence length $\xi$ of superconductors, as they are performed in high external magnetic fields. Other key parameters of superconductivity, such as the lower critical field $H_{c1}$ and London penetration depth $\lambda_L$, can be obtained from magnetic susceptibility measurements. This type of measurements is crucial for understanding the complex behavior of type-II superconductors in a magnetic field, particularly for the study of vortices.

Two techniques are currently used for magnetic susceptibility measurements at high pressures: a double modulation *ac* technique with a system of coils[27], and a superconducting quantum interference device (SQUID)[28]. The first technique provides only a qualitative indication of superconducting transition and is not widely distributed due to its complexity and low sensitivity[29,30]. Conversely, static *dc* measurements in SQUID are simpler, and the results are well-interpreted because they comprise absolute values of magnetic susceptibility. Initially, such measurements were limited to a pressure of approximately 10 GPa, and the registered signal was severely distorted by the magnetic background of bulky diamond anvil cell (DAC)[31,32]. The development of a miniature DAC capable of generating pressure values above 150 GPa[1,28] and recent improvements in the background subtraction procedure[33] have greatly extended the scope of the method. Nevertheless, these measurements are at the limit of their experimental capability.

Herein, we dramatically improved studies of the magnetic properties of superconductivity at high pressures by introducing the trapped flux method. This method is based on the detection of the magnetic flux, that originates from a conjunctive action of several properties/effects in superconductors (the persistent dissipation-free currents flow, quantum split of magnetic field into Abrikosov vortices, the pinning of the vortices by structural defects, etc.) and traps in a sample after switching off an external magnetic field. These measurements have several advantages over typical magnetic susceptibility measurements, as they are performed in zero magnetic field and thus, the huge magnetic background signal from the DAC is cancelled. The signal from the trapped flux can be approximately one-two orders higher than the response from the zero-field cooled (ZFC) field-screened state at low magnetic fields, and therefore, significantly smaller samples can be studied. This method is particularly beneficial for the study of superconductors having large pinning (which has been observed in several recently discovered superconductors), and is effective at distinguishing between different phases in multiphase samples, probing critical current densities $j_c$ in a wide temperature range, and is more precise at determining $H_{c1}$.

## RESULTS

$Im$-$3m$-$H_3S$ sample pressurized at $P$ ~155±5 GPa showed pronounced superconducting transition in ZFC $m(T)$ magnetization measurements with a $T_c$ ~195 K, 18 months after its synthesis (see Figure 1a). At lower temperatures, the sample underwent the second superconducting transition with a $T_c$ ~15 K (seen as a separation of the ZFC and field-cooled (FC) portions of $m(T)$ data on the background of the DAC signal in Figure 1b). The detected low-temperature superconductivity most probably corresponded to elemental sulfur, which was extruded on bevels of diamond anvils during the pressurization of the sandwiched sample of $NH_3BH_3$+S and remained intact after the pulsed laser-assisted synthesis (see inset in Figure 1b). The superconducting transition in S was not sharp and extended in the temperature range of ~12-18 K, which could be attributed to the substantial pressure gradients on the beveled slope of the diamond anvils. According to the universal diamond edge Raman scale[34], the pressure dramatically decreased from ~140 GPa at the edges of the anvil culet to ~95 GPa at the outer edge of the ring-shaped sample of sulfur. The measured $T_c$ was in good agreement with previous observations of superconductivity in sulfur at high pressures obtained from electrical transport[35] and $ac$ magnetic susceptibility[29] measurements.

**1. Trapped flux measurements.** The magnetization of a type-II superconductor is irreversible due to the pinning of vortices if the superconductor contains defects, such as dislocations, precipitates, and phase boundaries, which interact with the flux lines penetrating the superconductor above $H_{c1}$. We created a trapped flux in $Im$-$3m$-$H_3S$ under ZFC and FC conditions (run 1 and 2). ZFC mode consisted of cooling the sample at zero field from $T>T_c$ to the target magnetization temperature $T_M<T_c$, switching on an applied magnetic field $H_M$ and gradually decreasing of the magnetic field to 0 T (see details in Methods section). In FC mode, the sample was cooled from $T>T_c$ at applied magnetic field $H_M$, which was gradually decreased to 0 T at the target lowest temperature $T_M$.

The temperature-dependencies of the trapped magnetic moment $m_{trap}(T)$ generated in $H_3S$ under ZFC and FC conditions at different $H_M$ and measured between 10 and 250 K at zero field clearly showed a superconducting transition with a $T_c$ ~195 K (see Figure 1c, d), and further revealed a steep upturn below ~15 K, which originated from pure superconducting sulfur. The entry of magnetic flux into sulfur occurred at $\mu_0 H_M$ = 30 mT in ZFC mode and already at 0.5 mT in FC mode, which were the lowest applied magnetic fields in two runs. This indicated that the metallic sulfur belonged either to type-II superconductors or dirty type-I superconductors[36,37]. To exclude the contribution of sulfur to the measured magnetic response, only a portion of the $m_{trap}(T)$ data collected above 30 K was used to interpret the temperature-dependence of the trapped flux in $H_3S$.

Figure 2a shows the magnetic moment trapped in the sample under ZFC and FC conditions at different $H_M$ and measured at 30 K. In FC mode even weak applied magnetic fields as low as 0.5 mT resulted in positive non-zero magnetic response of the trapped flux in $H_3S$. In this mode the magnetic flux penetrated the sample in the normal state at $T>T_c$ and became trapped at $T<T_c$ due to very strong pinning (see details below).

Contrary to the FC mode, the ZFC magnetization provided valuable data for estimating $H_p$, when the applied magnetic field starts to penetrate the sample[38-40] (Figure 1c). No trapped flux in ZFC $H_3S$ was observed up to $\mu_0 H_M$ = 45 mT. In this range, the external magnetic field was completely repelled from the superconducting sample. We estimated $\mu_0 H_p(10 K)$ = 42±3 mT from the data measured at 45 mT ≤ $H_M$ ≤ 125 mT (see inset in Figure 2a).

At higher $H_M$, magnetic flux penetrated the superconductor from the outer edges of the disk-shaped sample and gradually propagated towards its center and filled the sample. The entry of magnetic flux led to a reduction in the Meissner currents, and being trapped to the appearance of superconducting current loops and a corresponding increase in the sample magnetization at zero field. The absence of steps in the $m_{trap}(H_M)$ data and smooth increase in $m_{trap}$ with the increase in $H_M$ implied the absence of a weak link network in the superconducting sample[41].

The observed trapped magnetic moment in the superconducting $Im$-$3m$-$H_3S$ phase saturated under ZFC and FC conditions at different values of $H_M$. We qualitatively interpreted this behavior of magnetization process in terms of the classical critical state Bean model[42,43], while considering the reversible part of magnetization in type-II superconductors[38,44]. According to the model, the flux trapping saturates if the applied magnetic field reaches the value $H_M = 2H^* + H_p$ in ZFC mode and $H_M = H^*$ in FC mode, where $H^*$ is the "full penetration field" (Figure 2b).

The trapped magnetic moment became saturated ($m_{trap}^s$) at $\mu_0 H_M \approx 1.7$ T under ZFC conditions and at $\mu_0 H_M \approx 0.8$ T under FC conditions, and did not vary with further increase in $\mu_0 H_M$ up to 6 T, which was the highest applied magnetic field in the experiment (see Figure 1c and 2a). Thus, $\mu_0 H^* = 800 \pm 50$ mT. The observed saturation is defined by the maximum value of the trapped magnetic field in the sample and corresponds to the critical current density $j_c$ that can be achieved in a superconductor.

We also estimated $H_{c1}$, which is proportional to $H_p$ using the relation $H_{c1} = \frac{1}{1-N} H_p$. The demagnetization factor $N$ was derived from the absolute value of $\Delta m$, which was the difference in ZFC $m(T)$ magnetization measurements between the normal and superconducting states. For the thin disk-shaped $Im\text{-}3m\text{-}H_3S$ sample having $d \sim 85$ μm and $h \sim 2.8$ μm (estimated lower and upper limits of $h$ were 2.1 μm and 3.1 μm), the demagnetization correction was ~8.5 (~7.7–11.4 for the estimated approximate limits of $h$, see details in Methods section). Thus, $\mu_0 H_{c1}(10\text{ K})$ was approximately 0.36 T (0.32–0.48 T).

The obtained value of $\mu_0 H_p \sim 42$ mT (and consequently, the retrieved value of $H_{c1}$) was noticeably lower than ~95 mT derived from the virgin curves of hysteretic magnetization loops in classical $m(H)$ measurements[28,33]. We considered the trapped flux method to be more sensitive to the determination of the onset of magnetic flux entry because a large linear contribution of the initial portion of $m(H)$ was cancelled. The advantage of this method for type-II superconductors having strong pinning was illustrated in ref[44]. Additionally, the signal of the trapped magnetic moment at zero field did not contain the field-dependent magnetic background arising from the DAC, which is unavoidable in typical $m(H)$ magnetization measurements. It should be noted that the presence of tiny, fragmented parts of the superconducting phase and the ragged edges of the bulky sintered sample (see photo of sample in the inset in Figure 1b) could influence the observed $H_p$. The contribution of these imperfections to the total measured magnetic moment was almost negligible in $m(H)$ experiments because the magnetic flux penetrated these areas at significantly low applied magnetic fields due to the extremely large demagnetization factor. In the trapped flux method these areas conversely accommodated magnetic flux at lower applied magnetic fields and provided magnetic response before the magnetic flux entered into the massive sample. We expected the true value of $H_p$ and the derived values of $\lambda_L$ and the Ginsburg-Landau parameter $\kappa = \frac{\lambda_L}{\xi}$ to lie approximately between the values observed in the two types of measurements. Using the data of electrical resistance measurements under high magnetic fields[45], which provided a coherence length $\xi(10\text{ K})$ of approximately 1.85 nm, we obtained $\lambda_L(10\text{ K})$ of approximately ~37 nm (31–40 nm within the estimated limits of $h$), and $\kappa(10\text{ K}) \sim 20$ (17–22). The corresponding values estimated from the $m(H)$ magnetization measurements[33] were $\lambda_L(0) \sim 22$ nm (18–23 nm) and $\kappa(0) \sim 12$ (10–13).

**2. Critical current density.** In addition to the determination of fundamental characteristics $T_c$ and $\lambda_L$, we deduced another important characteristics of a superconductor from measured $m_{trap}(\mu_0 H_M, T)$ data

– the critical current density over a wide temperature range. According to the Bean critical state model[42,43], a concentric screening current pattern having a current density $j_c$ averaged over a sample thickness creates a magnetic moment expressed as $m = \frac{\pi}{3} j_c h \left(\frac{d}{2}\right)^3$.

The measured value of $m_{trap}^s$ at 30 K corresponds to $j_c$(30 K) ~7.1(1)×10$^{10}$ A m$^{-2}$. The extrapolation to 0 K of the experimental data for $m_{trap}^s$ generated at $\mu_0 H_M$ = 2, 3, 4 and 6 T provides values of $m_{trap}^s(0)$ ~1.64(2)×10$^{-8}$ A m$^2$ and $j_c(0)$ ~7.3(1)×10$^{10}$ A m$^{-2}$. The calculated values of $j_c$ in H$_3$S exceed $j_c$(4.2 K) in single crystals of iron-based superconductors by two orders of magnitude[46,47] and are approximately two times lower than $j_c$(20 K) in REBCO coated conductors (2G-wires) of similar thickness[48].

The ultimate critical current density is limited by the depairing currents[49] as $j_d = \frac{\phi_0}{3\sqrt{3}\pi\mu_0} \frac{1}{\lambda_L^2 \xi}$, where $\phi_0$ is the magnetic flux quantum. Substituting the estimated values of $\lambda_L$ and $\xi$ we obtained $j_d$ ~4×10$^{13}$ A m$^{-2}$. The ratio between $j_c(0)$ and $j_d$ provided additional information. The values of $j_c(0)/j_d$ being approximately in the order of 10$^{-2}$-10$^{-1}$ is conventionally observed in low-temperature superconductors, which characterized by relatively low Ginzburg-Landau parameter $\kappa$ of ~20 and strong pinning that stems from the interaction of vortices with extended defects[50]. However, lower values of $j_c(0)/j_d$ of approximately 10$^{-3}$-10$^{-2}$ are characteristic for high-temperature superconductors, for example, the cuprate family in which the Ginzburg-Landau parameter is large ($\kappa$~100) and relatively weaker pinning that originated from point defects, such as oxygen vacancies[51]. To improve the in-field critical current performance of cuprates, different types of artificial pinning centers were introduced into the sample structure. The synthesized sample of H$_3$S was characterized by strong pinning (see below), but the ratio of $j_c(0)/j_d$ was quite small (approximately 10$^{-3}$). This suggested that the density of defect centers in the sintered sample was not maximal and could further be enhanced, for example, by modifying the synthesis protocol.

The different factors influencing the flux pinning in type-II superconductors are divided into two main types: $\delta T_c$ and $\delta l$[50,52]. The $\delta l$-type pinning arises from a spatial variation in the mean free path of charge carriers, and the defects are small and point sized, e.g. hydrogen vacancies in hydrides. The $\delta T_c$-type pinning is caused by a spatial variation of the Ginzburg-Landau parameter due to fluctuations in $T_c$, and the defects are larger than $\xi$, such as dislocations, grain boundaries and deviations of the chemical composition. Pressure gradients can also cause variations in $T_c$ in case of samples in DACs. We compared the temperature-dependence of the normalized critical current density $j_c(T)/j_c(0)$ in H$_3$S sample with the simulated curves for $\delta T_c$ and $\delta l$ types of pinning (see Figure 3). Both theoretical curves for pure $\delta T_c$ and $\delta l$ types of pinning deviate from experimental data; however, the resulting fit of the data by the combination of two models showed that the $\delta T_c$-type of pining dominates over the $\delta l$-type.

**3. Pinning and thermally activated motion of vortices.** The kinetics of the trapped magnetic moment demonstrates extremely slow rate of flux creep even at temperatures in vicinity of $T_c$, at which thermal fluctuations must be significantly higher (see Figure 4c). The estimated creep rate $S = \frac{1}{m_{trap,0}} \frac{dm_{trap}}{d\ln t}$ in *Im-3m*-H$_3$S was approximately 0.002 (or 500 ppm/h) at 165 K and approximately 0.005 (or 2000 ppm/h) at 185 K, which was comparable to the lowest values of $S$ measured in type-II superconductors[53], including the extremely slow creep rate in high-$j_c$ MgB$_2$ films at substantially lower $T/T_c < 0.5$[54]. The scale of fluctuations responsible for vortex melting[55] and vortex creep[53] in a superconductor can be quantified using the Ginzburg-Levanyuk number $Gi = \frac{1}{2}\left(\frac{2\pi\mu_0 k_B T_c \lambda_L^2}{\phi_0^2 \xi}\right)^2$, were $k_B$ is the Boltzmann constant. $Gi \sim 7\times 10^{-6}$ in H$_3$S was substantially lower than the values reported for cuprate ($\sim 10^{-2}$) and iron-based ($\sim 10^{-5}$–$10^{-2}$) high-temperature superconductors[53,55], and comparable to those reported in low-temperature superconductors ($\sim 10^{-9}$–$10^{-6}$) and MgB$_2$ ($\sim 10^{-6}$). The observed slow creep rate $S$ in H$_3$S was close to the theoretical limit demarcated by the approximate $Gi^{\frac{1}{2}}\frac{T}{T_c}$ line[53] and was in good agreement with the lower values of $\lambda_L$ and $\kappa$ (larger values of $\lambda_L$ would lead to the theoretical limit of $S$ being higher than the observed value). Additionally, electrical transport measurements of *Im-3m*-H$_3$S at high magnetic fields[45] showed weak vortex fluctuations and narrow vortex liquid regions due to the low values of $\kappa$ and $Gi$.

On account of strong pinning of vortices in H$_3$S sample the amount of the trapped flux solely depended on $H_M$ at which the flux was generated and the maximum temperature at which the sample was exposed after the magnetization step (see Figure 4). For example, the trapped magnetic moment created at $\mu_0 H_M = 1$ T by three different protocols merged at 100 K and above upon subsequent warming (brown and orange curves of cycle II, purple curve of cycle III, and green curve of cycle IV in Figure 4a). The behavior of the trapped magnetic moment was found to be temperature-independent if the sample was warmed to a certain temperature below $T_c$ and cooled down again (see horizontal trend of $m_{trap}(T)$ in Figure 4a and b). No pronounced hysteresis was observed for the magnetic moment during the cooling/warming cycles. This memory effect, which allows one to control the amount of trapped flux in the superconducting sample, stems from the very strong pinning of the vortices.

**4. Trapped flux in lanthanum superhydrides.** The trapped magnetic flux was further probed in the lanthanum-hydrogen system. The ZFC curve of $m(T)$ data measured at an applied magnetic field of 10 mT demonstrate an extended transition with the onset of $T_c \sim 200$ K at 120±5 GPa (Figure 5a). The observed $T_c$ was in good agreement with the values of the *C2/m*-LaH$_{10}$ phase at similar pressures measured by electrical transport measurements[56]. The transition at $T_c \sim 200$ K was significantly broader than that measured in the initial sample of *Fm-3m*-LaH$_{10}$ at 130±8 GPa with a $T_c \sim 231$ K[33], thus indicating the poorer quality of the superconducting phase. Likely, the high-temperature superconducting *Fm-3m*-LaH$_{10}$ phase sustained structural distortions into the *C2/m*-LaH$_{10}$ phase and partial decomposition into the hydrogen-deficient phases due to the unexpected drop in pressure during transportation from the synchrotron (see Methods section).

Contrarily, the trapped flux method demonstrates more pronounced superconducting transitions in this sample. In addition to the detection of the transition at ~200 K in the $C2/m$-LaH$_{10}$ phase, it revealed a superconducting state of another phase below ~70 K. Superconductivity with a $T_c$ ~70 K was previously observed in several samples prepared from La and H$_2$ (taken in a large deficiency) at approximately 150-178 GPa by four-probe electrical transport measurements and could be related to LaH$_x$ (where 3 < x < 10)[10]. We were unable to perform measurements of the trapped flux generated at different $H_M$ as we did for the sample of H$_3$S. At magnetization of the sample at $\mu_0 H_M$ = 4 T, one of the diamonds cracked, and pressure dropped below ~10 GPa. However, this allowed us to measure the background response of the DAC body induced after sweeping off a high magnetic field of 4 T. No remnant nonlinear magnetic background of the DAC body or other anomalies in the reference $m_{trap}(T)$ curve were detected in the DAC at ~10 GPa when the sample was evidently not superconducting (cycle III in Figure 5b). Furthermore, we clearly demonstrated that the observed superconducting transitions in the $m_{trap}(T)$ data collected at 120±5 GPa (cycles I and II in Figure 5b) belonged to the sample and not to the high-pressure assembly (DAC, gasket, diamonds, etc.).

**5. Discussion.** The Meissner effect is one of the most fundamental visualization of superconductivity and it has been considered as the proof for the bulk superconductivity. However, a magnitude of the Meissner effect in type-II superconductors substantially varies in experiments from almost complete expulsion in pinning-free conventional type-II superconductors[57,58], to practically no expulsion in iron pnictides[59], or even to magnetic moment enhancement in various materials with extreme sensitivity to disorder[60-63]. The specific features of the Meissner effect in high-temperature superconductors and its strong suppression by external magnetic fields in the mixed state are associated with the strong pinning which prevents the vortices inside the sample from shifting towards the sample surface even after crossing the $H_{c1}(T)$ line[64,65]. The subtle or scarcely observable Meissner effect (in FC mode) in H$_3$S[33] agrees with very strong pinning observed in the present work in H$_3$S even at temperatures near $T_c$. Under FC conditions, the strong pinning leads to the trapping of magnetic flux in the superconducting H$_3$S already at 0.5 mT, i.e. much lower than $H_p$. Such strong pinning of vortices may also be promoted by the high-pressure conditions. The absence of significant Meissner effect in elemental sulfur and LaH$_{10}$ supports this hypothesis, though systematic measurements of different high-pressure superconductors are required to prove it.

In contrast to standard $m(T)$ magnetization measurements at low magnetic fields, which demonstrate the shielding of an external magnetic field in the Meissner state, the registered value of the trapped magnetic moment is substantially larger (see Figure 1). The maximum value of a trapped magnetic moment $m_{trap}^s$ ~16.5×10$^{-9}$ A m$^2$ in $Im$-$3m$-H$_3$S at 20 K was approximately 40 times larger than the value of $\Delta m$ ~4.5×10$^{-10}$ A m$^2$ observed in ZFC $m(T)$ magnetization measurements at $\mu_0 H$ = 4 mT. Importantly, the response from the trapped flux at zero magnetic field was not perturbed by the magnetic background stemming from the bulky body of the DAC, including diamond anvils and a rhenium gasket. This significantly simplifies measurements of superconducting samples at high pressures in SQUID because the large background from the DAC is eliminated. Additionally, unlike electrical transport measurements, the trapped flux method is sensitive to a particular superconducting phase in mixtures because it does not require the specific arrangement of different phases relative to electrical leads. The analysis of the $m_{trap}(T)$ data provides estimations for important characteristics of superconductivity, such as $H_{c1}$, $\lambda_L$, pinning of vortices, and especially $j_c$, which can be probed in a wide temperature range. We believe that the trapped flux method, which shows great benefit for high-pressures studies, has also been underestimated for the study of superconductivity at ambient pressure. It can be a powerful tool for the routine screening of new superconducting materials, the study of multiphase, contaminated samples, or samples with a low superconducting fraction.

## MATERIALS AND METHODS

### Samples

We used the same samples, which were studied in our previous work[33]. The superconducting $Im\text{-}3m$-$H_3S$ and $Fm\text{-}3m$-$LaH_{10}$ phases were synthesized from sandwiched samples $S+NH_3BH_3$ and $LaH_3+NH_3BH_3$ pressurized in miniature DACs, which were specially designed for a standard commercial SQUID magnetometer[1,33]. The details of the preparation of DACs, chemical synthesis of the samples, estimation of pressure values, and characterization of the initial reactants and final products can be found in Ref.[33]

The DAC with $Im\text{-}3m$-$H_3S$ sample retained the same pressure value of $P = 155\pm5$ GPa throughout all measurements, whereas the DAC with $Fm\text{-}3m$-$LaH_{10}$ sample did not survive the transportation from APS synchrotron in Argonne, USA. The pressure decreased from $P = 130\pm8$ GPa to $P = 120\pm5$ GPa. We attempted to restore the sample and pressurized the DAC to $P = 125\pm5$ GPa and heated the sample to ~2000 K by a pulsed YAG laser. However, the sample demonstrated a lower $T_c$ ~200 K and likely corresponded to the structurally distorted $C2/m$ phase of $LaH_{10}$, which formed from $Fm\text{-}3m$-$LaH_{10}$ upon decompression at pressures below ~130 GPa[56]. One of the diamond anvils cracked, and the pressure dropped to ~10 GPa during the magnetization step at $\mu_0 H_M = 4$ T after two successful measurements of the trapped flux generated at 1 T and 2 T. The subsequent measurements of the DAC with the decomposed non-superconducting sample magnetized at $\mu_0 H_M = 4$ T demonstrated the linear background signal of the DAC with no anomalies in $m_{trap}(T)$ curve over the entire temperature range of 10-280 K.

### Magnetization measurements

Magnetization measurements $m(T)$ were performed using the S700X SQUID magnetometer by Cryogenic Limited, and a miniature DAC was attached to a 140-mm-long straw made of Kapton polyimide film, which was specially designed to minimize end effects. The position of the sample relative to the pickup coil of SQUID magnetometer was determined using the ferromagnetic signal from a small steel piece with a size of approximately 140×100×25 μm³ attached directly to the rhenium gasket surrounding the sample. The precision of the centering procedure is approximately ±0.2 mm. The superconducting transition in H$_3$S and LaH$_{10}$ was probed by ZFC and FC *m(T)* measurements at 4 mT and 10 mT, respectively. A $T_c$ ~195 K in the *Im-3m*-H$_3$S phase, which was determined as the offset of the diamagnetic transition on the ZFC curves of *m(T)*, was in good agreement with the values estimated in our previous measurements[33]. The much broader superconducting transition in the sample with LaH$_{10}$ and the lower $T_c$ ~200 K indicated the deterioration of the superconducting phase of LaH$_{10}$ after altering the pressure in the DAC, which likely sustained the monoclinic structural distortions[56].

The trapped flux was generated in two different protocols under ZFC (run 1) and FC (run 2) conditions. The typical ZFC protocol included cooling of the sample at zero field from $T>T_c$ to the desired temperature $T_M$ below the corresponding $T_c$s, namely $T_M$ = 10 K for *Im-3m*-H$_3$S sample and at $T_M$ = 4 K for *C2/m*-LaH$_{10}$ sample. The magnetic flux was also trapped in H$_3$S at $T_M$ = 4 K in cycle I, $T_M$ = 100 K in cycles II and IV, and $T_M$ = 8 K and 25 K in cycles V$^a$ and V$^b$. A typical magnetization cycle in ZFC mode consisted of a gradual increase in the applied magnetic field $H_M$ perpendicular to the sample surface at the lowest temperature point $T_M$, standing at the target value of $H_M$ for an hour, and a gradual decrease in the magnetic field to 0 T. In FC mode the sample was cooled from its normal state ($T>T_c$) at the applied magnetic field $H_M$ to the target temperature $T_M<T_c$. Then $H_M$ was gradually decreased to 0 T. After removing the applied magnetic field $H_M$ in both ZFC and FC modes the SQUID magnetometer was allowed to stand at 0 T for approximately 5 hours to prevent successive measurements of $m_{trap}(T)$ from outliers associated with jumps of flux in a superconducting magnet. An applied magnetic field of magnetization $\mu_0 H_M$ ranged from 30 mT to 6 T. $m_{trap}(T)$ measurements were performed upon warming (cooling) of the sample with a temperature step of 0.3-3 K (a smaller step in the vicinity of $T_c$) and 3-4 iterations at each temperature point. Additionally, the reference $m_{trap}(T)$ data were measured for *Im-3m*-H$_3$S sample by skipping the magnetization cycle ($\mu_0 H_M$ = 0 T). The trapped magnetic moment was determined as the difference between the measured magnetic moment after magnetization cycle and the residual magnetic moment, which arose from the body of the miniature DAC above the corresponding $T_c$ (see Supplementary Figure 1 and 2). Before each cycle of magnetization, the sample was converted to the normal state by warming to ambient temperature, and the superconducting magnet of SQUID was degaussed to eliminate the remaining fields.

Measurements of the trapped flux created in the sample of H$_3$S in run 1 and 2 were separated in time of ~6 months. The small discrepancy of the value of $m_{trap}^s$ between two runs likely aroused from the slightly different positions of the superconducting sample relative to the pickup coil of SQUID magnetometer, which were found at the centering procedure, and not from the different conditions of the creation of the trapped flux (FC or ZFC). The fact that the same values of $m_{trap}^s$ measured in run 1 under ZFC (2, 3 and 6 T) and FC (4 T) conditions support this explanation.

**Estimation of sample size and demagnetization correction**

The diameter and thickness of the thin disk-shaped *Im-3m*-$H_3S$ sample were estimated from optical microscopy and X-ray diffraction data as 85 μm³ and 2.8 μm³, respectively (theoretical lower and upper limits of *h* were 2.1 μm and 3.1 μm). Only the geometry of the ideal disk-shaped sample yielded high values of the demagnetization correction $\frac{1}{1-N}$ of approximately 20 using the proposed equation for the effective demagnetization factor in ref[66]. However, the real shape of sample, particularly the alteration of the thickness, can deviate from the ideal disk. Therefore, we further considered the absolute value of *Δm*, the difference in *m(T)* between a normal metal state (above $T_c$) and a superconducting state (below $T_c$), which includes geometrical imperfections in the sample shape. This approach gives more reasonable and reliable values of demagnetization correction ~8.5 (~7.7–11.4 for the estimated limits of *h*), which we use for the estimation of $H_{c1}$ (see details in Methods in ref[33]).

## ACKNOWLEDGEMENTS

M.I.E. is thankful to the Max Planck community for their support, and Prof. Dr. U. Pöschl for the constant encouragement. Work at the Ames Laboratory (S.L.B.) was supported by the U.S. Department of Energy, Office of Science, Basic Energy Sciences, Materials Sciences and Engineering Division under Contract No. DE-AC02-07CH11358. E.F.T. acknowledges financial support by the Ministry of Science and Higher Education of the Russian Federation through Grant No. AAAA-A18-118020190104-3 and through a Ural Federal University project within the Priority-2030 Program. The authors are thankful to Dr. V. G. Kogan and Prof. Dr. J. E. Hirsch for valuable discussions.

## AUTHOR CONTRIBUTIONS

V.S.M. and M.I.E. designed the research. M.I.E. designed the miniature diamond anvil cell. V.S.M prepared samples and performed measurements. V.S.M., V.K., S.L.B, and E.F.T. processed and analyzed the data. V.S.M. and M.I.E. wrote the manuscript with input from all co-authors.

## COMPETING INTERESTS

The authors declare no competing interests.

# FIGURES

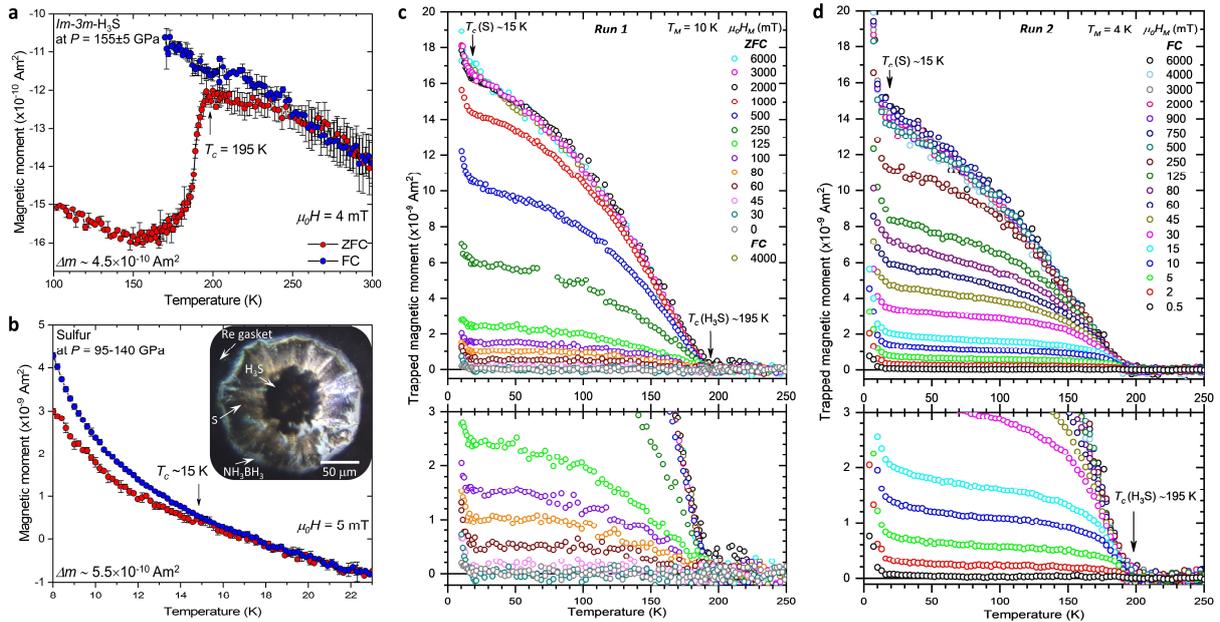

**Figure 1.** Magnetic measurements of sample containing H$_3$S and elemental sulfur. **a**) ZFC and FC $m(T)$ data of *Im-3m*-H$_3$S at 155±5 GPa. **b**) ZFC and FC $m(T)$ data of sulfur at 95-140 GPa, which remained intact on the diamond anvil bevels. The inset is a photo of the sample showing the spatial distribution of H$_3$S and S. **c**) and **d**) Temperature dependence of a trapped magnetic moment at zero field generated under ZFC (run 1) and FC (run 2) conditions, respectively. Open circles of different colors correspond to temperature-dependence of the trapped flux created at different magnetic fields $\mu_0 H_M$ (0 – 6 T). The lower panels reveal the beginning of penetration of the flux lines into H$_3$S sample above $\mu_0 H_M$ = 45 mT (run 1) and already at $\mu_0 H_M$ = 0.5 mT (run 2).

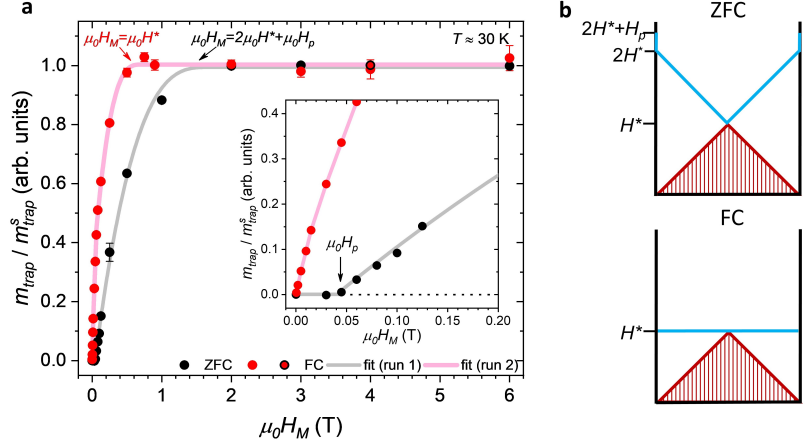

**Figure 2.** Trapped magnetic moment in *Im-3m*-$H_3S$ sample at 30 K. **a)** Dependence of a trapped magnetic moment at zero field on magnetization field $\mu_0 H_M$ measured in ZFC and FC modes (run 1 and 2). Circles correspond to the experimental data, magenta and grey curves are guides for the eye. Trapped flux was created in ZFC (black circles) and FC (red circle) modes at several applied magnetic fields $H_M$. The inset with the enlarged plot shows the entry of magnetic field into the sample at low $H_M$. **b)** The profile of magnetic field in the disk-shaped sample in applied magnetic field $H_M = 2H^* + H_p$ (ZFC mode) and $H_M = H^*$ (FC mode) (area below blue lines) and after removing the applied magnetic field (hatched red area).

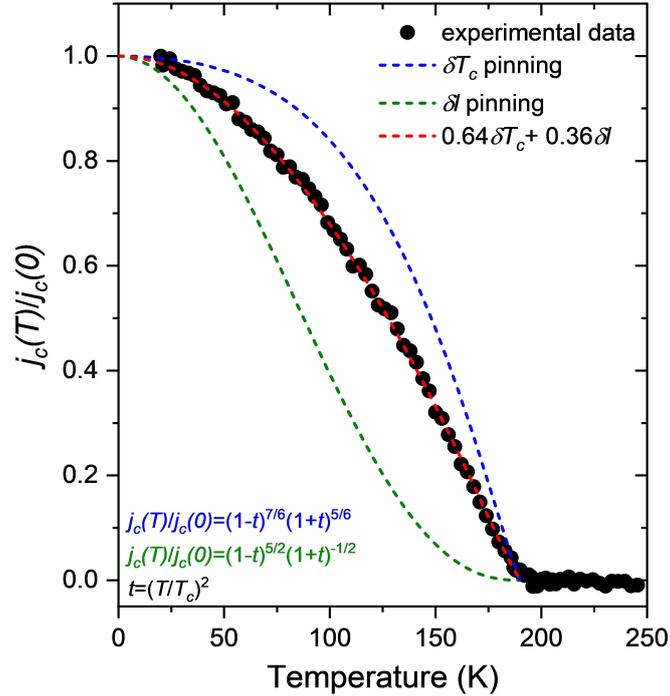

**Figure 3.** The temperature-dependence of the normalized zero field critical current density $j_c(T)/j_c(0)$ in *Im-3m*-$H_3S$ sample. Black circles are the experimental data (trapped flux is created at $T_M = 10$ K and $\mu_0 H_M = 3$ T). Dashed blue and green curves correspond to the temperature dependence of critical current density limited by $\delta T_c$- and $\delta l$-type pinning; red dashed curve is the fit of experimental data by $\delta T_c + \delta l$ model.

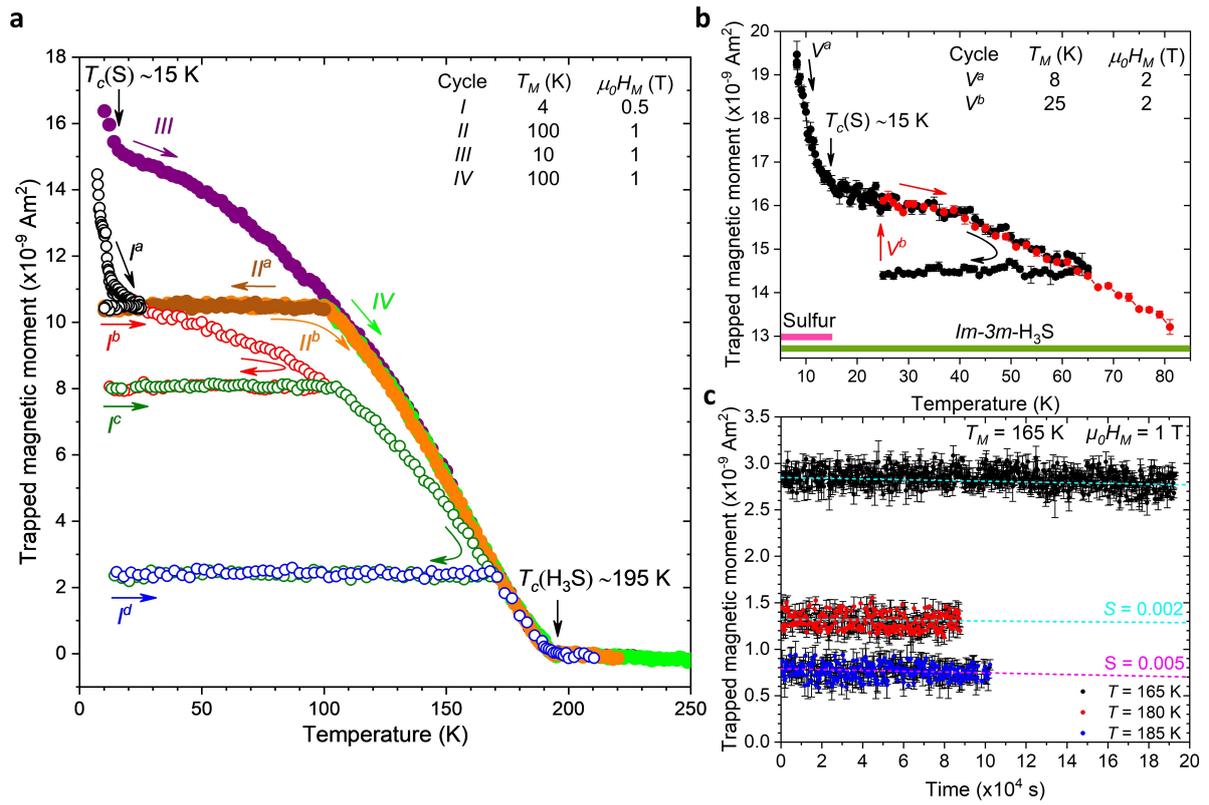

**Figure 4.** Vortex pinning in $H_3S$. **a**) Temperature-dependence of a trapped magnetic moment created at different temperatures $T_M$ and magnetic fields $H_M$ of magnetization. Arrows show the change of temperature in different warming/cooling cycles. **b**) Temperature-dependence of a trapped magnetic moment in $H_3S$ generated at $\mu_0 H_M$ = 2 T and $T_M$ = 8 K (black data, cycle $V^a$). The trapped flux can be restored if the sample is magnetized again (red data, cycle $V^b$). **c**) Creep of the trapped flux in $H_3S$ at several temperatures near a $T_c$ (the trapped flux was generated at $\mu_0 H_M$ = 1 T and $T_M$ = 165 K).

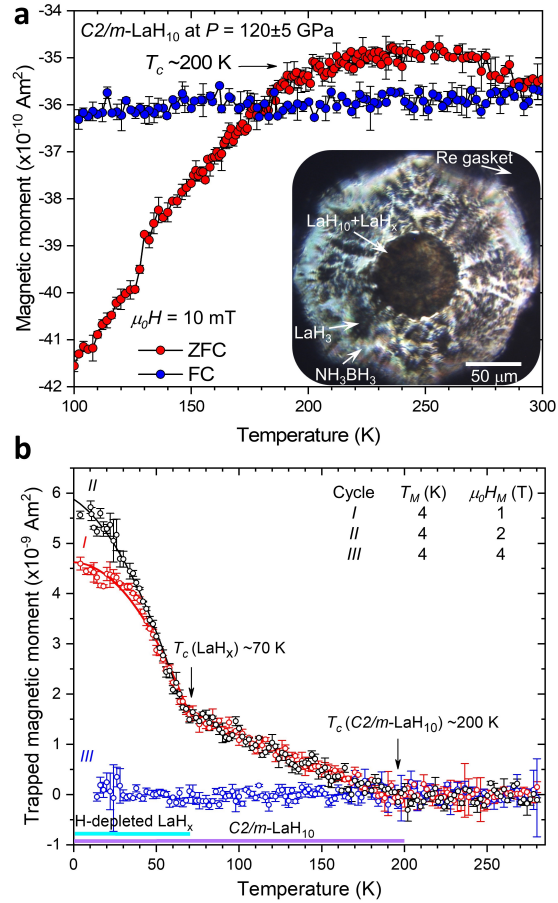

**Figure 5**. Magnetic measurements of sample containing $C2/m$-LaH$_{10}$ and LaH$_x$ (3 < x < 10). **a)** ZFC and FC $m(T)$ data of $C2/m$-LaH$_{10}$ and LaH$_x$ at 120±5 GPa. Photo of the sample is shown in the inset. **b)** Temperature-dependence of a trapped magnetic moment at zero field. Red and black open circles and fitting curves of different colors correspond to temperature-dependence of the trapped flux created under ZFC conditions at a magnetic field $\mu_0 H_M$ = 1 T and 2 T, respectively. Blue open circles are the data of the same sample at ~10 GPa after magnetization at $\mu_0 H_M$ = 4 T, showing the absence of a non-linear magnetic background from the DAC without superconducting sample.

# REFERENCES

1. Drozdov, A. P., Eremets, M. I., Troyan, I. A., Ksenofontov, V. & Shylin, S. I. Conventional superconductivity at 203 K at high pressures. *Nature* **525**, 73 (2015).
2. Drozdov, A. P. *et al.* Superconductivity at 250 K in lanthanum hydride under high pressures *Nature* **569**, 528 (2019).
3. Somayazulu, M. *et al.* Evidence for Superconductivity above 260 K in Lanthanum Superhydride at Megabar Pressures. *Phys. Rev. Lett.* **122**, 027001 (2019).
4. Schilling, A., Cantoni, M., Guo, J. & Ott, H. Superconductivity above 130 k in the Hg–Ba–Ca–Cu–O system. *Nature* **363**, 56-58 (1993).
5. Parkin, S. S. P. *et al.* Bulk Superconductivity at 125 K in $Tl_2Ca_2Ba_2Cu_3O_x$. *Physical Review Letters* **60**, 2539-2542, doi:10.1103/PhysRevLett.60.2539 (1988).
6. Bardeen, J., Cooper, L. N. & Schrieffer, J. R. Theory of Superconductivity. *Phys. Rev.* **108** 1175-1204 (1957).
7. Eliashberg, G. Interactions between electrons and lattice vibrations in a superconductor. *Sov. Phys. JETP* **11**, 696-702 (1960).
8. Migdal, A. Interaction between electrons and lattice vibrations in a normal metal. *Sov. Phys. JETP* **7**, 996-1001 (1958).
9. Kong, P. *et al.* Superconductivity up to 243 K in the yttrium-hydrogen system under high pressure. *Nature Communications* **12**, 5075, doi:10.1038/s41467-021-25372-2 (2021).
10. Drozdov, A. P. *et al.* Superconductivity at 250 K in lanthanum hydride under high pressures *Nature* **569** 528 (2019).
11. Somayazulu, M. *et al.* Evidence for Superconductivity above 260 K in Lanthanum Superhydride at Megabar Pressures. *Phys. Rev. Lett.* **122** 027001 (2019).
12. Hong, F. *et al.* Superconductivity of Lanthanum Superhydride Investigated Using the Standard Four-Probe Configuration under High Pressures. *Chin. Phys. Lett..* **37**, 107401 (2020).
13. Duan, D. *et al.* Pressure-induced metallization of dense (H2S)2H2 with high-Tc superconductivity. *Scientific Reports* **4**, 6968, doi:10.1038/srep06968 (2014).
14. Liu, H., Naumov, II, Hoffmann, R., Ashcroft, N. W. & Hemley, R. J. Potential high-Tc superconducting lanthanum and yttrium hydrides at high pressure. *Proc. Natl. Acad. Sci.* **114**, 6990 (2017).
15. Peng, F. *et al.* Hydrogen Clathrate Structures in Rare Earth Hydrides at High Pressures: Possible Route to Room-Temperature Superconductivity. *Phys. Rev. Lett.* **119** 107001 (2017).
16. Flores-Livas, J. A. *et al.* A perspective on conventional high-temperature superconductors at high pressure: Methods and materials. *Physics Reports* **856**, 1-78, doi:https://doi.org/10.1016/j.physrep.2020.02.003 (2020).
17. Boeri, L. *et al.* The 2021 room-temperature superconductivity roadmap. *Journal of Physics: Condensed Matter* **34**, 183002, doi:10.1088/1361-648x/ac2864 (2022).
18. Pickard, C. J., Errea, I. & Eremets, M. I. Superconducting Hydrides Under Pressure. *Annual Review of Condensed Matter Physics* **11**, 57-76, doi:10.1146/annurev-conmatphys-031218-013413 (2020).
19. Zhong, X. *et al.* Potential room temperature superconductivity in clathrate Lanthanide/Actinides octadechydrides at extreme pressures. *Research Square*, doi:https://doi.org/10.21203/rs.3.rs-1148583/v1 (2021).
20. Sun, Y., Lv, J., Xie, Y., Liu, H. & Ma, Y. Route to a Superconducting Phase above Room Temperature in Electron-Doped Hydride Compounds under High Pressure. *Physical Review Letters* **123**, 097001, doi:10.1103/PhysRevLett.123.097001 (2019).
21. Lucrezi, R., Di Cataldo, S., von der Linden, W., Boeri, L. & Heil, C. In-silico synthesis of lowest-pressure high-Tc ternary superhydrides. *npj Computational Materials* **8**, 119, doi:10.1038/s41524-022-00801-y (2022).
22. Di Cataldo, S., Heil, C., von der Linden, W. & Boeri, L. $LaBH_8$: Towards high-$T_c$ low-pressure superconductivity in ternary superhydrides. *Physical Review B* **104**, L020511, doi:10.1103/PhysRevB.104.L020511 (2021).
23. Bi, J. *et al.* Efficient route to achieve superconductivity improvement via substitutional La-Ce alloy superhydride at high pressure. *arXiv*, doi:10.48550/ARXIV.2204.04623 (2022).

| | |
|---|---|
| 24 | Schilling, J. S. Studies in superconductivity at extreme pressures. *Physica C* **460–462**, 182–185 (2007). |
| 25 | Hamlin, J. J. Superconductivity in the metallic elements at high pressures. *Physica C* **514** 59–76 (2015). |
| 26 | Shimizu, K. Superconductivity from insulating elements under high pressure. *Physica C* **514** 46–49 (2015). |
| 27 | Timofeev, Y. A. Detection of superconductivity in high-pressure diamond anvil cell by magnetic susceptibility technique. *Prib. Tekh. Eksper.* **5**, 186–189 (1992). |
| 28 | Eremets, M. I. *et al.* High-Temperature Superconductivity in Hydrides: Experimental Evidence and Details. *Journal of Superconductivity and Novel Magnetism*, doi:10.1007/s10948-022-06148-1 (2022). |
| 29 | Struzhkin, V. V., Hemley, R. J., Mao, H.-k. & Timofeev, Y. A. Superconductivity at 10–17 K in compressed sulphur. *Nature* **390**, 382-384, doi:10.1038/37074 (1997). |
| 30 | Struzhkin, V. *et al.* Superconductivity in La and Y hydrides: Remaining questions to experiment and theory. *Matter and Radiation at Extremes* **5**, 028201, doi:10.1063/1.5128736 (2020). |
| 31 | Alireza, P. L. & Lonzarich, G. G. Miniature anvil cell for high-pressure measurements in a commercial superconducting quantum interference device magnetometer. *Review of Scientific Instruments* **80**, 023906, doi:10.1063/1.3077303 (2009). |
| 32 | Marizy, A., Guigue, B., Occelli, F., Leridon, B. & Loubeyre, P. A symmetric miniature diamond anvil cell for magnetic measurements on dense hydrides in a SQUID magnetometer. *High Pressure Research* **37**, 465-474, doi:10.1080/08957959.2017.1387255 (2017). |
| 33 | Minkov, V. S. *et al.* Magnetic field screening in hydrogen-rich high-temperature superconductors. *Nat Commun* **13**, 3194, doi:10.1038/s41467-022-30782-x (2022). |
| 34 | Eremets, M. I. *et al.* Universal diamond edge Raman scale to 0.5 terapascal: The implication to metallization of hydrogen. *ArXiv*, doi:https://doi.org/10.48550/arXiv.2202.06933 (2022). |
| 35 | Kometani, S. *et al.* Observation of Pressure-Induced Superconductivity of Sulfur. *Journal of the Physical Society of Japan* **66**, 2564-2565, doi:10.1143/JPSJ.66.2564 (1997). |
| 36 | Treimer, W., Ebrahimi, O., Karakas, N. & Prozorov, R. Polarized neutron imaging and three-dimensional calculation of magnetic flux trapping in bulk of superconductors. *Physical Review B* **85**, 184522, doi:10.1103/PhysRevB.85.184522 (2012). |
| 37 | Dhiman, I. *et al.* Role of Temperature on Flux Trap Behavior in < 100 > Pb Cylindrical Sample: Polarized Neutron Radiography Investigation. *Physics Procedia* **69**, 420-426, doi:https://doi.org/10.1016/j.phpro.2015.07.059 (2015). |
| 38 | Altshuler, E., García, S. G. & Barroso, J. Flux trapping in transport measurements of YBa2Cu3O7-x superconductors. *Physica C-superconductivity and Its Applications* **177**, 61-66 (1991). |
| 39 | Buntar, V., Sauerzopf, F. M. & Weber, H. W. Lower critical fields of alkali-metal-doped fullerene superconductors. *Physical Review B* **54**, R9651-R9654, doi:10.1103/PhysRevB.54.R9651 (1996). |
| 40 | Buntar, V. & Weber, H. W. Magnetic properties of fullerene superconductors. *Superconductor Science and Technology* **9**, 599-615, doi:10.1088/0953-2048/9/8/001 (1996). |
| 41 | Müller, K. H., Andrikidis, C., Du, J., Leslie, K. E. & Foley, C. P. Connectivity and limitation of critical current in Bi-Pb-Sr-Ca-Cu/Ag tapes. *Physical Review B* **60**, 659-666, doi:10.1103/PhysRevB.60.659 (1999). |
| 42 | Bean, C. P. Magnetization of Hard Superconductors. *Physical Review Letters* **8**, 250-253, doi:10.1103/PhysRevLett.8.250 (1962). |
| 43 | Bean, C. P. Magnetization of High-Field Superconductors. *Reviews of Modern Physics* **36**, 31-39, doi:10.1103/RevModPhys.36.31 (1964). |
| 44 | Moshchalkov, V. V., Henry, J. Y., Marin, C., Rossat-Mignod, J. & Jacquot, J. F. Anisotropy of the first critical field and critical current in YBa2Cu3O6.9 single crystals. *Physica C: Superconductivity* **175**, 407-418, doi:https://doi.org/10.1016/0921-4534(91)90616-7 (1991). |
| 45 | Mozaffari, S. *et al.* Superconducting phase diagram of $H_3S$ under high magnetic fields. *Nature Communications* **10**, 2522 (2019). |

46  Yamamoto, A. *et al.* Small anisotropy, weak thermal fluctuations, and high field superconductivity in Co-doped iron pnictide Ba(Fe1−xCox)2As2. *Applied Physics Letters* **94**, 062511, doi:10.1063/1.3081455 (2009).

47  Sun, Y. *et al.* Deviation from Canonical Collective Creep Behavior in Li0.8Fe0.2OHFeSe. *Journal of the Physical Society of Japan* **88**, 034703, doi:10.7566/JPSJ.88.034703 (2019).

48  Molodyk, A. *et al.* Development and large volume production of extremely high current density YBa2Cu3O7 superconducting wires for fusion. *Scientific Reports* **11**, 2084, doi:10.1038/s41598-021-81559-z (2021).

49  Griessen, R. *et al.* Evidence for mean free path fluctuation induced pinning in $YBa_2Cu_3O_7$ and $YBa_2Cu_4O_8$ films. *Physical Review Letters* **72**, 1910-1913, doi:10.1103/PhysRevLett.72.1910 (1994).

50  Blatter, G., Feigel'man, M. V., Geshkenbein, V. B., Larkin, A. I. & Vinokur, V. M. Vortices in high-temperature superconductors. *Reviews of Modern Physics* **66**, 1125-1388, doi:10.1103/RevModPhys.66.1125 (1994).

51  Van Gennep, D., Hassan, A., Luo, H. & Abdel-Hafiez, M. Sharp peak of the critical current density in $BaFe_{2-x}Ni_xAs_2$ at optimal composition. *Physical Review B* **101**, 235163, doi:10.1103/PhysRevB.101.235163 (2020).

52  Dew-Hughes, D. Flux pinning mechanisms in type II superconductors. *The Philosophical Magazine: A Journal of Theoretical Experimental and Applied Physics* **30**, 293-305, doi:10.1080/14786439808206556 (1974).

53  Eley, S., Miura, M., Maiorov, B. & Civale, L. Universal lower limit on vortex creep in superconductors. *Nature Materials* **16**, 409-413, doi:10.1038/nmat4840 (2017).

54  Thompson, J. R. *et al.* Vortex pinning and slow creep in high-$J_c$ MgB2 thin films: a magnetic and transport study. *Superconductor Science and Technology* **18**, 970-976, doi:10.1088/0953-2048/18/7/008 (2005).

55  Koshelev, A. E. *et al.* Melting of vortex lattice in the magnetic superconductor $RbEuFe_4As_4$. *Physical Review B* **100**, 094518, doi:10.1103/PhysRevB.100.094518 (2019).

56  Sun, D. *et al.* High-temperature superconductivity on the verge of a structural instability in lanthanum superhydride. *Nature Communications* **12**, 6863, doi:10.1038/s41467-021-26706-w (2021).

57  Huebener, R. *Magnetic Flux Structures of Superconductors*.  (Springer, 2001).

58  Poole, C. P. J., Farach, H. A., Creswick, R. J. & Prozorov, R. *Superconductivity*. 3rd Edition edn,  870 (Elsevier, 2014).

59  Prozorov, R. *et al.* Anomalous Meissner effect in pnictide superconductors. *Physical Review B* **82**, 180513, doi:10.1103/PhysRevB.82.180513 (2010).

60  Braunisch, W. *et al.* Paramagnetic Meissner effect in Bi high-temperature superconductors. *Physical Review Letters* **68**, 1908-1911, doi:10.1103/PhysRevLett.68.1908 (1992).

61  Braunisch, W. *et al.* Paramagnetic Meissner effect in high-temperature superconductors. *Physical Review B* **48**, 4030-4042, doi:10.1103/PhysRevB.48.4030 (1993).

62  Thompson, D. J., Minhaj, M. S. M., Wenger, L. E. & Chen, J. T. Observation of Paramagnetic Meissner Effect in Niobium Disks. *Physical Review Letters* **75**, 529-532, doi:10.1103/PhysRevLett.75.529 (1995).

63  Geim, A. K., Dubonos, S. V., Lok, J. G. S., Henini, M. & Maan, J. C. Paramagnetic Meissner effect in small superconductors. *Nature* **396**, 144-146, doi:10.1038/24110 (1998).

64  Tomioka, Y., Naito, M. & Kitazawa, K. The Meissner and shielding effects in niobium in relation to oxide superconductors. *Physica C: Superconductivity* **215**, 297-304, doi:https://doi.org/10.1016/0921-4534(93)90229-J (1993).

65  Moshchalkov, V. V. & Zhukov, A. A. The Meissner effect in superconductors with strong vortex pinning. *Physica B: Condensed Matter* **169**, 601-602, doi:https://doi.org/10.1016/0921-4526(91)90346-G (1991).

66  Prozorov, R. & Kogan, V. G. Effective Demagnetizing Factors of Diamagnetic Samples of Various Shapes. *Physical Review Applied* **10**, 014030, doi:10.1103/PhysRevApplied.10.014030 (2018).

# Supplementary Information for

# Magnetic flux trapping in hydrogen-rich high-temperature superconductors

V. S. Minkov[1]*, V. Ksenofontov[1], S. L. Bud'ko[2,3], E. F. Talantsev[4,5], M. I. Eremets[1]*

[1]*Max Planck Institute for Chemistry; Hahn-Meitner-Weg 1, 55128 Mainz, Germany*

[2]*Ames Laboratory, U.S. Department of Energy, Iowa State University; Ames, IA 50011, United States*

[3]*Department of Physics and Astronomy, Iowa State University; Ames, IA 50011, United States*

[4]*M.N. Mikheev Institute of Metal Physics, Ural Branch of the Russian Academy of Sciences, S. Kovalevskoy St 18, 620108 Ekaterinburg, Russian Federation*

[5]*NANOTECH Centre, Ural Federal University, Mira St 19, 620002 Ekaterinburg, Russian Federation*

*Corresponding authors. Email: v.minkov@mpic.de (V.S.M); m.eremets@mpic.de (M.I.E.)

**This PDF file includes:**

    Supplementary Figure S1-2

# SUPPLEMENTARY FIGURE

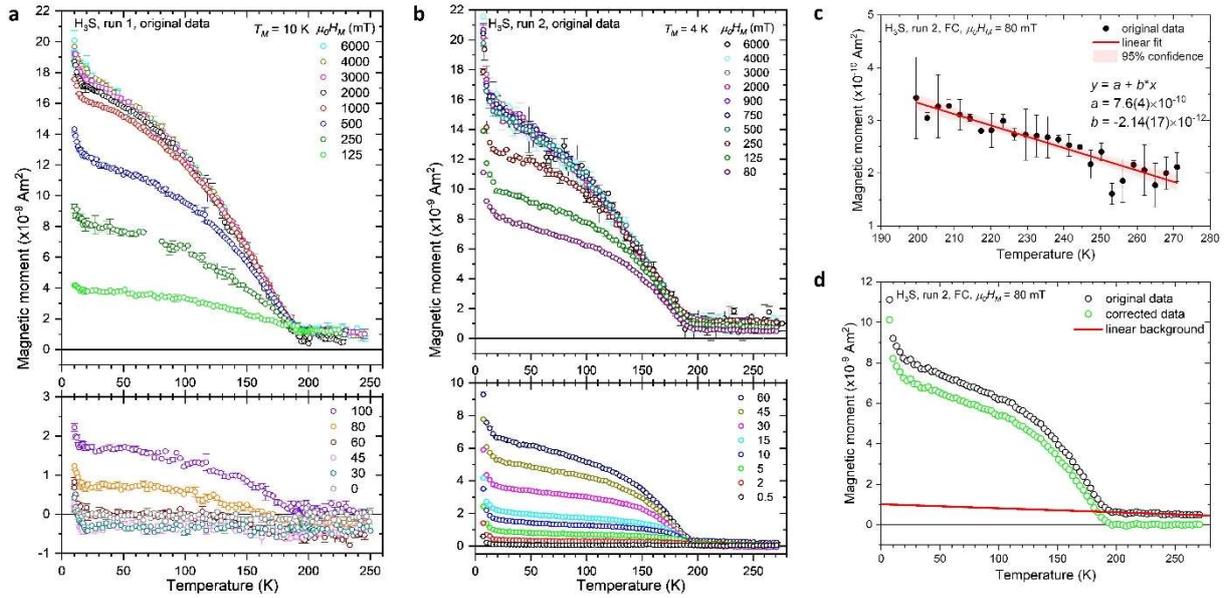

**Supplementary Figure S1.**

Raw temperature-dependent data of the trapped flux in the sample containing *Im-3m*-$H_3S$ and subtraction of the background. **a)** and **b)** Raw data of *m(T)* measured for *Im-3m*-$H_3S$ sample in run 1 and 2, respectively. **c)** The linear fit of the data above $T_c$ (red line) was defined as the background of the corresponding measurement. **d)** Example of the background subtraction for determination of the trapped magnetic moment in the superconducting sample. Black circles, red line and green circles correspond to original raw data, extrapolated linear background, and corrected $m_{trap}(T)$ for sample of $H_3S$ measured in run 2 at $\mu_0 H_M$ = 80 mT.

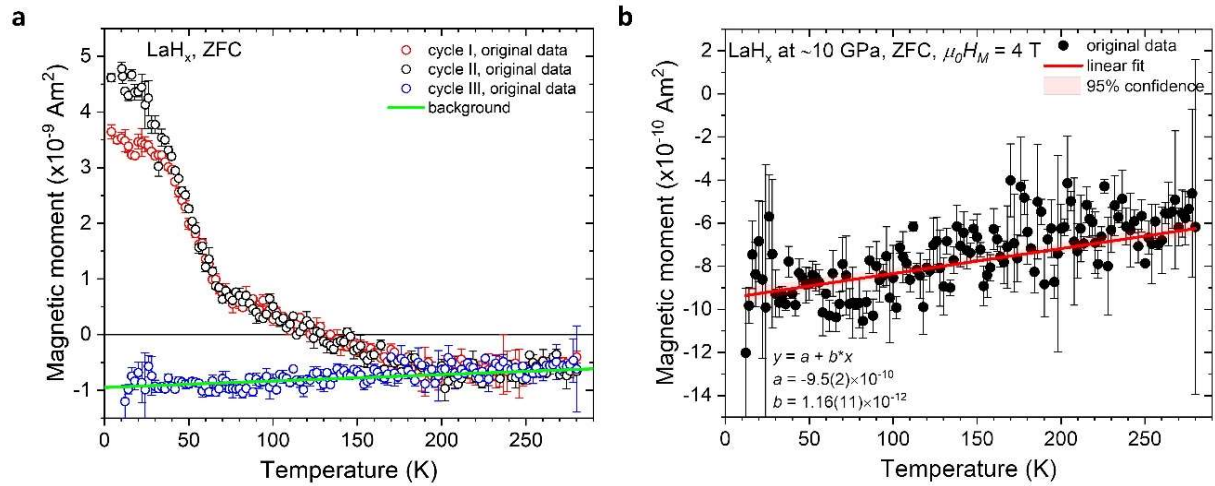

**Supplementary Figure S2.**

Subtraction of the background in raw temperature-dependent data of the trapped flux in sample containing $C2/m$-LaH$_{10}$. **a)** Raw data of $m(T)$ measured for sample with LaH$_{10}$. The trapped flux was generated under ZFC conditions at $T_M$ = 4 K and $\mu_0 H_M$ = 1 T and 2 T (red and black circles, respectively). **b)** The background was defined as a linear fit of the $m(T)$ data obtained for the evidently non-superconducting sample in DAC at ~10 GPa, which was magnetized under ZFC conditions at $\mu_0 H_M$ = 4 T after the decrease of pressure (blue circles).